\documentclass[sigplan,screen]{acmart}

\usepackage{main}
\usepackage{balance}

\setcopyright{cc}
\setcctype{by}
\acmDOI{10.1145/3779031.3779085}
\acmYear{2026}
\copyrightyear{2026}
\acmISBN{979-8-4007-2341-4/2026/01}
\acmConference[CPP '26]{Proceedings of the 15th ACM SIGPLAN International Conference on Certified Programs and Proofs}{January 12--13, 2026}{Rennes, France}
\acmBooktitle{Proceedings of the 15th ACM SIGPLAN International Conference on Certified Programs and Proofs (CPP '26), January 12--13, 2026, Rennes, France}
\received{2025-09-12}
\received[accepted]{2025-11-13}

\begin{document}

\title{Mechanizing Synthetic Tait Computability in Istari}

\author{Runming Li}
\orcid{0000-0001-7600-9069}
\affiliation{%
  \institution{Carnegie Mellon University}
  \city{Pittsburgh}
  \country{USA}
}
\email{runmingl@cs.cmu.edu}

\author{Yue Yao}
\orcid{0000-0001-8523-5156}
\affiliation{%
  \institution{Carnegie Mellon University}
  \city{Pittsburgh}
  \country{USA}
}
\email{yueyao@cs.cmu.edu}

\author{Robert Harper}
\orcid{0000-0002-9400-2941}
\affiliation{%
  \institution{Carnegie Mellon University}
  \city{Pittsburgh}
  \country{USA}
}
\email{rwh@cs.cmu.edu}

\begin{abstract}
  Categorical gluing is a powerful technique for proving meta-theorems of type theories such as canonicity and normalization. Synthetic Tait Computability (STC) provides an abstract treatment of the complex gluing models by internalizing the gluing category into a modal dependent type theory with a phase distinction. This work presents a mechanization of STC in the \istari{} proof assistant. \istari{} is a Martin-L\"{o}f-style extensional type theory with equality reflection, which avoids much of the explicit transport reasoning typically found in intensional proof assistants. This work develops a reusable library for synthetic phase distinction, including modalities, extension types, and strict glue types, and applies it to two case studies: (1) a canonicity model for dependent type theory with dependent products and booleans with large elimination, and (2) a Kripke canonicity model for the cost-aware logical framework. Our results demonstrate that the core STC constructions can be formalized essentially verbatim in \istari, preserving the elegance of the on-paper arguments while ensuring machine-checked correctness.
\end{abstract}

\keywords{gluing, synthetic Tait computability, logical relations, extensional type theory, \istari, equality reflection, meta-theory, cost-aware logical framework}

\ccsdesc[500]{Theory of computation~Type theory}
\ccsdesc[500]{Theory of computation~Categorical semantics}

\maketitle

\section{Introduction}\label{sec:intro}
The past decade has seen significant advancements in the meta-theory of programming languages, particularly for dependent type theories, due to the use of the categorical \emph{gluing} technique~\cite{crole>1994} in programming languages. Traditionally, meta-theorems of programming languages such as canonicity and normalization are proved using syntactic \emph{logical relations} arguments \ala Tait~\cite{tait>1967}. These are families of predicates or relations defined by induction on the structure of types, with respect to an operational semantics or a reduction system. When the syntax of a programming language is in more semantic and algebraic presentations, such as locally cartesian closed categories for dependent type theories, the gluing technique provides a categorical and \emph{proof-relevant} generalization to syntactic logical relations. Concretely, the gluing technique constructs a \emph{gluing} model over the syntactic model of the programming language, along a suitable functor to a semantic category, such as the category of sets. For example, gluing along the global sections functor yields a proof-relevant logical relations model that establishes canonicity~\cite{kaposi-huber-sattler>2019,bocquet-kaposi-sattler>2023,mitchell-scedrov>1992}, the property that all closed terms of boolean type are either $\true$ or $\false$. This technique has been applied to prove various canonicity and normalization results, including simply-typed $\lambda$-calculus~\cite{fiore>2002, sterling-spitters>2018}, System F~\cite{altenkirch-hofmann-streicher>1996}, dependent type theory~\cite{coquand>2018,kovacs>2026,altenkirch-chamoun-kaposi-shulman>2024}, and univalent type theory~\cite{shulman>2015}.

\subsection{Gluing}\label{sec:gluing-overview}
This subsection provides a brief overview of the gluing technique. 
For reasons of space, we assume familiarity with basic notions from category theory such as hom-functors and the Yoneda embedding, and refer the reader to classic textbooks such as~\citet{awodey>cattheory,riehl>cattheory}. 
Consider a category $\T$ that represents the syntax of an object language: its objects are types, and its morphisms are judgmental equivalence classes of terms. In this setting, a canonicity theorem may be formulated as follows:
\begin{theorem}[Canonicity]
  For every closed term of boolean type, represented as a morphism $b : \red{\terminal}_\T \to \red{\bool}_\T$ in $\T$, either $b = \red{\true}_\T$ or $b = \red{\false}_\T$.
\end{theorem}
To ensure the category representing the syntax has good categorical properties, we can embed $\T$ into presheaf category $\pre{\T}$ via the Yoneda embedding: $\Yo : \T \hookrightarrow \pre{\T}$. 
The proof strategy is to construct a model of the object language in the Artin gluing~\cite{sga4>1983} $\G = \set \downarrow \Gamma$, the comma category over the global sections functor:
\begin{align*}
    \Gamma & : \pre{\T} \to \set \\
    \Gamma & (A) = \homset{\Yo(\red{\terminal}_\T)}{A}
\end{align*}
under the identification $\Yo(\red{\terminal}_\T) = \terminal_{\pre{\T}}$ because Yoneda preserves limits.
To be explicit, an object in $\G$ is a triple $(S, A, f)$ where $S \in \set$, $A \in \pre{\T}$, and $f : S \to \Gamma(A)$; a morphism from $(S, A, f)$ to $(S', A', f')$ is a pair of morphisms $(h : S \to S', t : A \to A')$ such that the diagram as shown below commutes. An object in $\G$ in the form of $(S, \Yo(B), f)$ is to be thought of as a family of sets $S_b$ indexed by morphisms $\Yo(b) : \Yo(\red{\terminal}_\T) \to \Yo(B)$, \ie a proof-relevant predicate on closed terms of type $B$.
\[\begin{tikzcd}[sep=small]
	S &&& {S'} \\
	\\
	\\
	{\Gamma(A)} &&& {\Gamma(A')}
	\arrow["h", from=1-1, to=1-4]
	\arrow["f"', from=1-1, to=4-1]
	\arrow["{f'}", from=1-4, to=4-4]
	\arrow["{\Gamma(t)}", from=4-1, to=4-4]
\end{tikzcd}\]

A functorial model of the object language in $\G$ in the sense of~\citet{lawvere>thesis} is a functor $\iota : \pre{\T} \to \G$ that preserves necessary structures, such as finite products and exponential objects for simply-typed $\lambda$-calculus. An analogue of the fundamental theorem of logical relations in this setting is that the functorial model $\iota$ is a section of the projection functor $\pi : \G \to \pre{\T}$, \ie $\pi \circ \iota = \mathrm{Id}_{\pre{\T}}$ as depicted below. The action of $\pi$ on morphism $(h, t)$ is $\pi(h, t) = t : A \to A'$.
\[\begin{tikzcd}[sep=small]
	\T &&&& {\pre{\T}} &&&& {\mathcal{G}} \\
	\\
	\\
	&&&&&&&& \pre{\T}
	\arrow["\Yo", from=1-1, to=1-5]
	\arrow["\iota", from=1-5, to=1-9]
	\arrow["{\mathrm{Id}}"', equals, from=1-5, to=4-9]
	\arrow["\pi", from=1-9, to=4-9]
\end{tikzcd}\]

Suppose the action of $\iota$ on the boolean type $\iota(\Yo(\red{\bool}_\T)) = \blue{\BOOL}_\G$ is defined as 
\[
    \blue{\BOOL}_\G = \Presheaf[\begin{array}{lr}
        f(0) = \Yo(\red{\false}_\T) \\
        f(1) = \Yo(\red{\true}_\T)
    \end{array}]{\{0 , 1 \}}{\Gamma(\Yo(\red{\bool}_\T))}.
\]
Take any morphism $b : \red{\terminal}_\T \to \red{\bool}_\T$, which represents a closed term of boolean type in the object language. By the definition of $\iota$, the morphism $\iota(\Yo(b)) : \blue{\terminal}_\G \to \blue{\BOOL}_\G$ in the gluing category $\G$ contains the following data $(h : \terminal_{\set} \to \{0,1\}, t : \Yo(\red{\terminal}_{{\T}}) \to \Yo(\red{\bool}_\T))$:
\[\begin{tikzcd}[sep=small]
	{\terminal_{\set}} &&& {\{0, 1\}} \\
	\\
	\\
	{\Gamma(\Yo(\red{\terminal}_\T)) = \terminal_{\set}} &&& {\Gamma(\Yo(\red{\bool}_\T))}
	\arrow["h", from=1-1, to=1-4]
	\arrow[from=1-1, to=4-1]
	\arrow["\begin{array}{c} f(0) = \Yo(\red{\false}_\T) \\f(1) = \Yo(\red{\true}_\T) \end{array}", from=1-4, to=4-4]
	\arrow["{\Gamma(t)}", from=4-1, to=4-4]
\end{tikzcd}\]
From $\pi \circ \iota = \mathrm{Id}$ it follows that $\pi(\iota(\Yo(b))) = \Yo(b)$, and hence $t = \Yo(b)$. Considering the function $h : \terminal_{\set} \to \{0,1\}$, one obtains that $h$ maps to either $0$ or $1$. Consequently, $\Yo(b)$ must be either $\Yo(\red{\true}_\T)$ or $\Yo(\red{\false}_\T)$. By the full faithfulness of the Yoneda embedding, it follows that $b$ is either $\red{\true}_\T$ or $\red{\false}_\T$, thus establishing the canonicity theorem.

The heart of this proof is then to close the construction of $\iota$ under other type formers and term constructors of the object language, and check that $\iota$ is indeed a section of $\pi$ in each case. For a tutorial construction of common type formers in the gluing category, we refer readers to~\citet[\S 6.6]{angiuli-gratzer>2025}.

Formally, this gluing proof already works if we glue directly over the syntactic category $\T$ and the global sections functor $\Gamma : \T \to \set$. However, in anticipation of the type-theoretic reasoning in \cref{sec:stc}, working in the presheaf category $\pre{\T}$ allows us to leverage the internal language of presheaf categories to reason about the gluing construction in a type-theoretic manner.

\subsection{Synthetic Tait Computability}\label{sec:stc-intro}
The aforementioned gluing technique is a powerful and flexible tool for proving meta-theoretic properties of programming languages, but the construction of the gluing model can be quite involved (especially for those that prove normalization), with many tedious details to check. For example, depending on the exact presentation of the syntactic category $\T$, there may be many subtle but boring naturality conditions to verify. 

Sterling's \emph{synthetic Tait computability} (STC)~\cite{sterling>thesis} is a recent development that aims to simplify the construction of gluing models by internalizing the gluing construction in a suitable modal dependent type theory. This technique has been successfully applied to a variety of type theories, including modal~\cite{gratzer>2022} and cubical~\cite{sterling-angiuli>2021} type theories, ML-like module calculus~\cite{sterling-harper>2021}, dependent type theory with controlled unfolding~\cite{gratzer-sterling-angiuli-coquand-birkedal>2025}, higher-order effect handlers~\cite{yang>thesis,yang-wu>2026}, and dependent call-by-push-value~\cite{li-harper>2025}. 

The key idea of STC is to introduce a \emph{synthetic phase distinction} between \emph{\red{syntax}} and \emph{\blue{semantics}}. The idea of phase distinctions is originally due to the development of ML modules~\cite{harper-mitchell-moggi>1989} in which a \emph{static} phase isolates the static, compile-time constructs from dynamic, runtime constructs, where dynamic constructs can depend on static ones but not vice versa. The meta-theory of programming languages displays similar structures: the \emph{semantics} depends on the \emph{syntax}, but not vice versa, and a \emph{syntactic} phase can be used to isolate the syntactic constructs from the semantic construction. This isolation of syntax is formally analogous to the projection functor $\pi : \G \to \pre{\T}$ in the gluing construction, which forgets the semantic information. The innovation of STC is to view the internal language of the gluing category $\G$ as a modal extensional dependent type theory with a \emph{syntactic} phase, in which one can write down the definitions of the gluing model such as $\blue{\BOOL}_\G$ and the fact that $\pi(\blue{\BOOL}_\G) = \Yo(\red{\bool}_\T)$ using type-theoretic constructs, without having to explicitly construct the objects and morphisms in the gluing category. 

\subsubsection*{Synthetic Phase Distinction}
Synthetically in type theory, a \emph{syntactic} phase is an intuitionistic, type-theoretic proposition \syn{} that, when assumed in the context, isolates the syntax from the semantics. 
Categorically \syn{} may be interpreted as a subterminal object $(\initial_{\set}, \Yo(\red{\terminal}_\T), !)$ in $\G$,
which, when assumed in the context, erases the semantic component from $\set$ and leaves only the syntactic component from $\T$. A proposition like \syn{} immediately induces a pair of idempotent monadic modalities~\cite{rijke-shulman-spitters>2020}. 

\subsubsection*{Open Modality}
The open modality $\open{A} = \syn{} \to A$ is the reader monad for the phase \syn{} with monadic unit $\eta_{\open{}} : A \to \open{A}$ defined as $\eta_{\open{}} = \lambda a . \lambda z . a$. Intuitively, open modality introduces a syntactic assumption to the context, thereby isolating the syntax. Formally there is an equivalence of categories between $\T$ and the slice category $\G/{\syn{}}$, where the projection functor $\pi$ is equivalent to exponentiating by $\syn{}$. The condition that $\pi(\blue{\BOOL}_\G) = \Yo(\red{\bool}_\T)$ can then be expressed in the internal language as an open equation $\open(\blue{\BOOL} = \red{\bool})$. 

\subsubsection*{Closed Modality}
The closed modality $\closed{A} = A \lor \syn{}$ is the join between $A$ and $\syn{}$, which can be defined categorically as the pushout along the projection maps of $A \times \syn{}$, or equivalently a quotient inductive type that equates all elements under the syntactic phase as follows:
\begin{center}
  \begin{minipage}{0.4\linewidth}
    \[\begin{tikzcd} {A \times \syn{}} & \syn{} \\
        A & {\closed A}
        \arrow["{\pi_1}", from=1-1, to=2-1]
        \arrow["{\pi_2}", from=1-1, to=1-2]
        \arrow["{\star}", from=1-2, to=2-2]
        \arrow["{\eta_{\closed{}}}", from=2-1, to=2-2]
        \arrow["\lrcorner"{anchor=center, pos=0.125, rotate=180}, draw=none, from=2-2, to=1-1]
      \end{tikzcd}\]
  \end{minipage}%
  \begin{minipage}{0.6\linewidth}
    \iblock{
      \mhang{\kw{data}~\closed{}~(A : \univ) : \univ~\kw{where}}{
        \mrow{\Label{\eta_{\closed{}}} : A \to \closed A}
        \mrow{\Label{\star} : \syn{} \to \closed A}
        \mrow{\Label{law} : (a : A)~(z : \syn{}) \to \Label{\eta_{\closed{}}}~a = \Label{\star}~z}
      }
    }
  \end{minipage}
\end{center}
Intuitively, closed modality marks a type as purely semantic by identifying all its elements under the syntactic phase: $\open{\closed{A}}$ is contractible, \ie has exactly one element up to equality.
Categorically in $\G$, the construction of open and closed modalities can be understood as an exponential object and a pushout respectively as follows.
\[
    \open{} \Presheaf[f]{S}{\Gamma(A)} = \Presheaf[\mathsf{id}]{\Gamma(A)}{\Gamma(A)} \qquad \closed{} \Presheaf[f]{S}{\Gamma(A)} = \Presheaf[!]{S}{\terminal_{\set}}
\]

\subsubsection*{Other Dependent Type Formers}
The Artin gluing $\G = \set \downarrow \Gamma$ is an elementary topos~\cite{johnstone>topostheory}, meaning that the internal language of $\G$ supports common dependent type formers, such as dependent products and sums, binary coproducts, extensional equality types, and a hierarchy of universes. We refer readers to~\citet[\S 5.3.3]{yang>thesis} for an exposition of other type formers in $\G$.

\subsection{Formalization of Synthetic Tait Computability}\label{sec:formalization-intro}
This work addresses the formalization of synthetic Tait computability in a proof assistant. The STC approach to meta-theory is particularly attractive for mechanization: the internal language of the gluing category $\G$ is a dependent type theory that closely resembles the languages of many existing dependently-typed proof assistants. By extending a given proof assistant with the modal constructions required for STC, the definitions and proofs of STC can be expressed directly within that system. The on-paper proofs using STC typically deal with a large number of equalities coming from both the syntax of the object language and the meta-level modality framework. Nevertheless, such arguments are concise and elegant: for example, the canonicity proof for a core dependent type theory occupies less than a page. 
The following key factors contribute to this concision, and the subsequent discussion explains how they are addressed in the formalization of this work.

\subsubsection*{Equations in the Syntax} 
In an algebraic presentation, the equational theory of the object language is typically encoded as propositional equalities. Because the language of STC has extensional equality types, these propositional equalities can be internalized as judgmental equalities using \emph{equality reflection}. Therefore reasoning about equalities is hidden in the background (as part of the typing derivation). In a proof assistant like \agda{} where equality reflection is not available, these propositional equalities must be explicitly transported in the proofs, leading to large terms with transports and equalities between those terms that obscure the main ideas of the proofs. While there are efforts to turn limited forms of propositional equalities into judgmental ones via rewrite rules in \agda{}~\cite{cockx>2019}, this approach is not as general as equality reflection. For this reason, the present mechanization is carried out in \istari~\cite{istarilogic}, a Martin-L\"{o}f-style~\cite{martin-lof>1982} extensional proof assistant with a computational semantics, which provides equality reflection natively.

\subsubsection*{Phase Distinction}
A notable feature of synthetic phase distinction is that a term could have different types depending on the phase. For example, since $\open(\blue{\BOOL} = \red{\bool})$, any term $b : \blue{\BOOL}$ is also $b : \red{\bool}$ under the syntactic phase. This kind of implicit coercion is used heavily as a convenient device in on-paper STC proofs. The logic of \istari is a type-assignment system in the sense of \nuprl, in which a term may be assigned different types. This allows the implicit coercions to be used exactly as in on-paper STC proofs.

\subsubsection*{Extension Types}
In STC, proof obligations like $\open(\blue{\BOOL} = \red{\bool})$ are achieved by using \emph{extension types}~\cite{riehl-shulman>2017}, $\extTy{A}{\syn{}}{a_0}$, the collection of elements of $A$ that restrict to $a_0$ under the syntactic phase \syn{}. 
Extension types are not natively supported in most proof assistants, but one option is to encode them as $\Sigma$-types $\Sigma_{a : A} \open(a = a_0)$. 
The $\Sigma$-type encoding, however, does not provide implicit coercion that is heavily used in on-paper STC proofs: if $a : \extTy{A}{\syn{}}{a_0}$ then $a : A$. On the other hand, extension types align well with subset types~\cite{allen-bickford-constable-eaton-kreitz-lorigo-moran>2006,nordstrom-petersson-smith>1990}, which \istari{} supports natively and gives the desired implicit coercion.

\subsection{Contributions}
This work presents a formalization of synthetic Tait computability in the \istari{} proof assistant, with the following contributions:
\begin{enumerate}[leftmargin=*,nosep]
    \item A library of synthetic phase distinction in \istari, including modalities, extension types, strict glue types, and other constructs necessary for STC;
    \item Formalizations of two STC case studies in \istari:
    \begin{enumerate}[leftmargin=*,nosep]
        \item A canonicity gluing model for a dependent type theory with a base type of booleans supporting large elimination and dependent products~\cite[\S 4.4]{sterling>thesis}, corresponding to unary logical relations;
        \item A canonicity gluing model for the cost-aware logical framework~\cite{li-harper>2025, niu-sterling-grodin-harper>2022}, a dependent call-by-push-value~\cite{levy>2003} language with a phase distinction for cost analysis, corresponding to unary Kripke logical relations~\cite{kripke>1963}.
    \end{enumerate}
\end{enumerate}

\subsubsection*{Synopsis}
The remainder of the paper is organized as follows. \Cref{sec:stc} provides a brief refresher on an example proof using synthetic Tait computability. \Cref{sec:istari} presents a tutorial on the \istari{} proof assistant and its underlying type theory. \Cref{sec:mech} introduces the library for synthetic phase distinction and illustrates the mechanization of STC through two case studies. \Cref{sec:related} discusses related work on the formalization of gluing arguments, and \cref{sec:future} outlines possible directions for future research.

\section{A Refresher on STC}\label{sec:stc}
This section provides a brief refresher on synthetic Tait computability by considering canonicity for a core dependent type theory with booleans, one of the case studies mechanized in \cref{sec:mech}. The discussion begins with an introduction of the technical devices required for STC.

\subsection{Extension Types}
Originally developed in the context of homotopy type theory~\cite{riehl-shulman>2017} and cubical type theory~\cite{redtt>2018}, where the ``phase'' is dimension formula for cubes, extension types $\extTy{A}{\syn}{a_0}$ classify the elements of a type $A$ that are equal to a distinguished element $a_0$ under the influence of $\syn$. In the context of STC, this construction provides a succinct and elegant formulation of the condition $\pi \circ \iota = \mathrm{Id}_\T$. 
Implicit coercions are employed to simplify notation: if $a : \extTy{A}{\syn}{a_0}$ then $a : A$ and $\open(a = a_0)$, and conversely. The standard inference rules for extension types are presented in \cref{fig:extension-rules}.

\begin{figure}[ht]
    \centering
    \begin{mathpar}
    \inferrule[Formation]
    {   A : \univ \\
        \syn{} \to (a_0 : A)
    }
    {\extTy{A}{\syn}{a_0} : \univ}
    \hspace{1.5em}
    \mprset{ fraction ={===}}
    \inferrule[Introduction \& Elimination]
    {a : A \\
        \syn{} \to (a = a_0 : A)
    }
    {a : \extTy{A}{\syn}{a_0}}
\end{mathpar}
    \caption{Inference rules for extension types}
    \label{fig:extension-rules}
\end{figure}

\subsection{Strict Glue Types}
The core idea of STC invites the existence of a strict glue type $\glue{a}{A}{B(a)}$, where a syntactic component $A$ is glued to a semantic component $B$ just like a $\Sigma$ type. The key difference is that it needs to be governed by the syntactic phase, so that open equations of the kind $\open(\glue{a}{A}{B(a)} = A)$ hold. In order for such equations to hold, the glue type needs to have the syntactic component $A$ be \emph{open-modal} and the semantic component $B$ be \emph{closed-modal}~\cite{rijke-shulman-spitters>2020}.  

\begin{definition}
    A type $A$ is \emph{open-modal} if its interpretation in the gluing category is a constant function, \ie an object in the form of $(\Gamma(A), A, \mathsf{id})$. In the internal language, this means $A$ is isomorphic to $\open A$: $A \cong \open A$.
\end{definition}

\begin{definition}
    A type $B$ is \emph{closed-modal} if its interpretation in the gluing category is trivial on the open part, \ie an object in the form of $(B, \Yo(\red{\terminal}_\T), !)$. In the internal language, this means $B$ is isomorphic to $\closed B$: $B \cong \closed B$. The present mechanization uses an equivalent definition~\cite{rijke-shulman-spitters>2020} that is easier to work with: $B$ is closed-modal if under $\syn$, the type $B$ is contractible, \ie $\open B \cong \terminal$. Immediately $\closed B$ for any type $B$ is closed-modal; $\extTy{A}{\syn}{a_0}$ is also closed-modal because syntactically it collapses to a singleton $a_0$.
\end{definition}
In the gluing category, the glue type $\glue{a}{A}{B(a)}$ is interpreted exactly as \emph{gluing} the interpretations of $A$ and $B$.
\[
A = \Presheaf[\mathsf{id}]{\Gamma(A)}{\Gamma(A)} \quad B(a) = \Presheaf[!]{B(a)}{\Gamma(\Yo(\red{\terminal}_\T))=\terminal_{\set}} 
\]
\[
\glue{a}{A}{B(a)} = \Presheaf[\pi_1]{\coprod_{a \in \Gamma(A)} B(a)}{\Gamma(A)}
\]
Notationally we write $\glueEx{\syn{}}{a}{b}$ for an element of the glue type $\glue{a}{A}{B(a)}$, equipped with projections $\pi_{\circ}$ and $\pi_{\bullet}$. The associated $\beta$- and $\eta$-equations hold as expected. The most significant equations are
\[
    \open (\glue{a}{A}{B(a)} = A) \quad \text{and} \quad \open (\glueEx{\syn{}}{a}{b} = a).
\]
These open equations are critical for the fundamental theorems of logical relations, where the open modality plays the role of projecting out the syntactic part. A complete set of inference rules for the strict glue type is presented in \cref{fig:sglue-rules}. The use of strict glue types in STC is standard~\cite{yang>thesis,sterling-harper>2022,sterling>2022-logical-relations,yang-wu>2026,gratzer-sterling-angiuli-coquand-birkedal>2025} and can be justified by the realignment/strictification axiom in a Grothendieck topos~\cite{birkedal-bizjak-clouston-grathwohl-spitters-vezzosi>2016,orton-pitts>2016,sterling-harper>2022,sterling>thesis,gratzer-shulman-sterling>2024}. Unlike \citet{sterling-harper>2021} who use realignment directly to construct computability argument, this work axiomatizes strict glue types to emphasize the geometric intuition of the syntax/semantics distinction. Conceptually, this is closely related to the glue types employed in other forms of phase distinction~\cite{grodin-li-harper>2025}, where the corresponding open equations are justified by univalence.

\begin{figure}[ht]
    \centering
    {\allowdisplaybreaks
\begin{mathpar}
    \inferrule[Formation \& Type-eq-syn]
    {   A : \syn{} \to \univ \\
        B : ((z : \syn{}) \to A~z) \to \univ \\
        (x : ((z : \syn{}) \to A~z)) \to (B(x) \cong \terminal)
    }
    {\glue{x}{A}{B(x)} : \univ \\ 
    (z : \syn{}) \to \glue{x}{A}{B(x)} = A~z : \univ}

    \inferrule[Introduction]
    {a : (z : \syn) \to A~z \\
        b : B(a)
    }
    {\glueEx{\syn{}}{a}{b} : \glue{x}{A}{B(x)}}

    \inferrule[Elimination-open \& Elimination-closed]
    {g : \glue{x}{A}{B(x)}}
    {\pi_\circ ~g : (z : \syn{}) \to A~z \\ \pi_\bullet ~g : B(\pi_\circ ~g)}

    \inferrule[Uniqueness]
    {g : \glue{x}{A}{B(x)}}
    {g = \glueEx{\syn{}}{\pi_\circ ~g}{\pi_\bullet ~g} : \glue{x}{A}{B(x)}}

    \inferrule[Computation-open \& Computation-closed]
    {a : (z : \syn) \to A~z \\
        b : B(a)
    }
    {\pi_\circ ~\glueEx{\syn{}}{a}{b} = a : (z : \syn{}) \to A~z \\
    \pi_\bullet ~\glueEx{\syn{}}{a}{b} = b : B(a)}

    \inferrule[Term-eq-syn]
    {
      g : \glue{x}{A}{B(x)}
    }
    {(z : \syn{}) \to (\pi_\circ ~g)~z = g : A~z}
\end{mathpar}
}
    \caption{Inference rules for strict glue types}
    \label{fig:sglue-rules}
\end{figure}

\subsection{Syntax}
The presentation of a type theory can be given succinctly as a signature in a logical framework following the \emph{judgments as types} principle~\cite{harper-honsell-plotkin>1993,sterling>thesis}. For example, the signature of a dependent type theory with booleans, large elimination, and dependent products is given below in a semantic logical framework~\cite{harper>2021,uemura>thesis,yang>2025,sterling>thesis}, that is, in the internal language of locally Cartesian closed categories (LCCC). The adequacy of this presentation is ensured by \citet{gratzer-sterling>2021} on defining dependent type theories in LCCCs. The syntactic category $\T$ is then the free LCCC generated by the constants of the signature, quotiented by the equalities between constants. By construction, all syntax is open-modal; in the mechanization this is enforced by taking the signature of the syntax under the \open{} modality.

This second-order algebraic presentation of syntax in the sense of Martin L\"{o}f's logical framework~\cite{harper>2021,kaposi-xie>2024,nordstrom-petersson-smith>1990} is particularly convenient and ergonomic for mechanization, in direct contrast to the inductive characterizations of syntax, which often require lengthy and intricate reasoning about bindings and substitutions as exemplified in many prior mechanization efforts for programming language meta-theory~\cite{poplmark,poplmark-reloaded}, as well as the use of quotients, which are not available in many proof assistants, to represent judgmental equality. 
{\allowdisplaybreaks
\begin{align*}                                                            
  \red{\tp}                 & : \univ \\ 
  \red{\tm}                 & : \red{\tp} \to \univ \\
  \red{\bool}               & : \red{\tp} \\
  \red{\true}               & : \red{\tm}(\red{\bool}) \\
  \red{\false}              & : \red{\tm}(\red{\bool}) \\
  \red{\ifelim}             & : (C : \red{\tm}(\red{\bool}) \to \red{\tp}) \to (b : \red{\tm}(\red{\bool})) \to \\ & \quad \red{\tm}(C (\red{\true})) \to \red{\tm}(C (\red{\false})) \to \red{\tm}(C(b)) \\
  \red{\ifelim_{\beta_1}}   & : \red{\ifelim}~C~\red{\true}~t~f = t \\
  \red{\ifelim_{\beta_2}}   & : \red{\ifelim}~C~\red{\false}~t~f = f \\ 
  \red{\pitp}               & : (A : \red{\tp}) \to (\red{\tm}(A) \to \red{\tp}) \to \red{\tp} \\
  \red{\lam}                & : ((x : \red{\tm}(A)) \to \red{\tm}(B(x))) \to \red{\tm}(\red{\pitp}~A~B) \\
  \red{\app}                & : \red{\tm}(\red{\pitp}~A~B) \to (x : \red{\tm}(A)) \to \red{\tm}(B(x)) \\
  \red{\pitp_\beta}         & : \red{\app}~(\red{\lam}~f)~ a = f~a \\
  \red{\pitp_\eta}          & : \red{\lam}~(\red{\app}~e) = e
\end{align*}
}

\subsection{Canonicity Model}
In specifying the functorial semantics $\iota : \pre{\T} \to \G$, the main task is to define the images of the constants produced by the functor $\iota$. More precisely, the goal is to define:
\begin{align*}
    \blue{\TP}             & : \extTy{\univ}{\syn}{\red{\tp}} \\
    \blue{\TM}             & : \extTy{\blue{\TP} \to \univ}{\syn}{\red{\tm}} \\
    \blue{\BOOL}           & : \extTy{\blue{\TP}}{\syn}{\red{\bool}} \\
    \blue{\TRUE}           & : \extTy{\blue{\TM}(\blue{\BOOL})}{\syn}{\red{\true}} \\
    \cdots
\end{align*}

\subsubsection*{Notation}
Throughout this paper, lowercase \red{\textsf{red}} terms denote syntactic components, whereas uppercase \blue{\textsf{BLUE}} terms denote their semantic counterparts under the image of $\iota$.

\subsubsection{Semantics of Judgments}\label{sec:stc-judgments}
The semantics of $\red{\tp}$ is given by the collection of all syntactic types in the language, each equipped with the corresponding collection of semantics of terms of that type, \cf the proof-relevant logical relations definition for universes~\cite{shulman>2015,coquand>2018}. The glue type is used to assemble all data into a single structure.\footnote{For simplicity, universe levels are omitted in the presentation, although they are fully accounted for in the formalization.} 

\iblock{
    \mrow{\Label{\blue{\TP}} : \extTy{\univ}{\syn}{\red{\tp}}}
    \mrow{\Label{\blue{\TP}} = \glue{A}{\red{\tp}}{\extTy{\univ}{\syn}{\red{\tm}(A)}}}
}
\noindent This definition satisfies the extension type condition $\open(\blue{\TP} = \red{\tp})$ by the equation of glue type. The semantics of $\red{\tm}$ is then obtained by projecting the corresponding term collection for each type $A : \blue{\TP}$.

\iblock{
    \mrow{\Label{\blue{\TM}} : \extTy{\blue{\TP} \to \univ}{\syn}{\red{\tm}}}
    \mrow{\Label{\blue{\TM}}~A = \pi_\bullet A}
}
\noindent This definition also satisfies the extension type condition $\open(\blue{\TM} = \red{\tm})$ by the use of extension type in $\blue{\TP}$.

\subsubsection{Semantics of Booleans}
In proving canonicity, the goal is to show that every closed term of boolean type is either $\red{\true}$ or $\red{\false}$. To this end, the semantics of $\red{\bool}$ is defined by gluing the syntactic boolean terms with a semantic component that classifies the canonical boolean values.

\iblock{
    \mrow{\Label{\blue{\BOOL}} : \extTy{\blue{\TP}}{\syn}{\red{\bool}}}
    \mhang{\Label{\blue{\BOOL}} = [ \syn{} \hookrightarrow \red{\bool} \mid }{
        \mrow{\glue{b}{\red{\tm}(\red{\bool})}{\closed(b = \red{\true} + b = \red{\false})} ] }
    }
}

\noindent To type-check this definition, three conditions need to be verified:
\begin{enumerate}[leftmargin=*,nosep]
    \item First, $\open(\blue{\BOOL} = \red{\bool})$ is required, which follows directly from the judgmental equations of terms of glue types.
    \item Second, the semantic component of $\blue{\BOOL}$ restricts to $\red{\tm}(\red{\bool})$ under $\syn$ by the glue type equation, as required by $\blue{\TP}$.
    \item Third, the use of the closed modality in the predicate $\closed(b = \red{\true} + b = \red{\false})$ guarantees that the semantic component of the glue type is closed-modal, as required by its formation rule.
\end{enumerate}

The semantics of terms of type $\red{\bool}$ are injections, and the semantics of $\red{\ifelim}$ is a case analysis on the disjunction in the semantic part of $\blue{\BOOL}$.

\iblock{
    \mrow{\Label{\blue{\TRUE}} : \extTy{\blue{\TM}(\blue{\BOOL})}{\syn}{\red{\true}}}
    \mrow{\Label{\blue{\TRUE}} =  \glueEx{\syn{}}{\red{\true}}{\eta_{\closed{}}(\mathsf{inl}(\checkmark))} }
    \mrow{}
}

\iblock{
    \mrow{\Label{\blue{\IF}} : \extTy{(C : \blue{\TM}(\blue{\BOOL}) \to \blue{\TP}) \to (b : \blue{\TM}(\blue{\BOOL})) \to \blue{\TM}(C(\blue{\TRUE})) \to \blue{\TM}(C(\blue{\FALSE})) \to \blue{\TM}(C(b))}{\syn}{\red{\ifelim}}}
    \mhang{\Label{\blue{\IF}}~C~b~t~f = \mathsf{case}~ \pi_\bullet b~ \mathsf{of}}{
        \mrow{\eta_{\closed{}}(\mathsf{inl}(\_)) \Rightarrow t}
        \mrow{\eta_{\closed{}}(\mathsf{inr}(\_)) \Rightarrow f}
        \mrow{\star~z \Rightarrow \red{\ifelim}~C~b~t~f}
    }
}
\noindent Note that non-trivial equality proof obligations in the definition of $\blue{\IF}$ arise in order to ensure that the three branches coincide, as the construction defines a map out of a quotient.

\subsubsection{Semantics of Dependent Products}
A recurring pattern emerges in the definition of these constants: the heart of the proof is expressed concisely in type-theoretic terms (sometimes called the \emph{realizers}), followed by accompanying English explanations (the typing derivations) to justify the non-trivial well-typedness conditions. Because the internal language is extensional, reasoning about equalities occurs at the judgmental level rather than in the surface syntax. This observation motivates the mechanization of STC in an extensional proof assistant: the definitions should remain concise as-is, and the type-checker can ensure that all well-typedness conditions are formally satisfied. In other words, in extensional type theory, the proof consists of not only the term and type, but also its typing derivation. For instance, if the mechanization accepts the following definition of $\blue{\PITP}$, then all necessary conditions can be guaranteed.

\iblock{
    \mrow{\Label{\blue{\PITP}} : \extTy{(A : \blue{\TP}) \to (\blue{\TM}(A) \to \blue{\TP}) \to \blue{\TP}}{\syn}{\red{\pitp}}}
    \mhang{\Label{\blue{\PITP}}~A~B = [ \syn{} \hookrightarrow \red{\pitp}~A~B \mid }{
        \mrow{\glue{e}{\red{\tm}(\red{\pitp}~A~B)}{\extTy{(a : \blue{\TM}(A)) \to \blue{\TM}(B(a))}{\syn}{\red{\app}~e}} ] }
    }
    \mrow{}
    \mrow{\Label{\blue{\LAM}} : \extTy{((x : \blue{\TM}(A)) \to \blue{\TM}(B(x))) \to \blue{\TM}(\blue{\PITP}~A~B)}{\syn}{\red{\lam}}}
    \mrow{\Label{\blue{\LAM}}~f = \glueEx{\syn{}}{\red{\lam}~f}{f} }
    \mrow{}
    \mrow{\Label{\blue{\APP}} : \extTy{\blue{\TM}(\blue{\PITP}~A~B) \to (x : \blue{\TM}(A)) \to \blue{\TM}(B(x))}{\syn}{\red{\app}}}
    \mrow{\Label{\blue{\APP}}~e~a = (\pi_\bullet e)~a }
}

It is worth noting that, for the definition of $\blue{\LAM}$ to be well-typed, the equation $\red{\pitp_\beta} : \red{\app}~ (\red{\lam}~f)~ a = f~a$ from the syntax must be used. This equation does not appear explicitly in the term because, again, the internal language is extensional; by equality reflection, it is turned into a judgmental equality and can thus be used directly in type-checking.
The $\beta$- and $\eta$-equations must also be verified, which can be done straightforwardly using the equations for glue types.

\iblock{
    \mrow{\Label{\blue{\PITP_\beta}} : \extTy{\blue{\APP}~ (\blue{\LAM}~f)~ a = f~a}{\syn}{\red{\pitp_\beta}}}
    \mrow{\Label{\blue{\PITP_\beta}} = \checkmark}
    \mrow{}
    \mrow{\Label{\blue{\PITP_\eta}} : \extTy{\blue{\LAM}~(\blue{\APP}~e) = e}{\syn}{\red{\pitp_\eta}}}
    \mrow{\Label{\blue{\PITP_\eta}} = \checkmark}
}

\subsection{What Is So Synthetic About STC?}\label{sec:synthetic}
Compared to \emph{analytical} mathematics, \emph{synthetic} methods emphasize the use of high-level axiomatizations and abstractions to express ideas directly, rather than constructing them from more primitive notions. It is important to highlight the key axiomatizations that STC uses and how they impact the mechanization.

\begin{axiom}\label{ax:phase-classifier}
    There is an intuitionistic proposition $\syn{}$ that classifies the syntactic phase.
\end{axiom}

\Cref{ax:phase-classifier} is crucial in STC to isolate and simplify the categorical formulation of the fundamental theorem of logical relations: $\pi \circ \iota = \mathrm{Id}_\T$ becomes type-theoretic extension-type conditions such as $\blue{\BOOL} : \extTy{\blue{\TP}}{\syn}{\red{\bool}}$. In the mechanization, this axiom can be realized by postulating such a proposition and its associated properties, as is done in other formalizations of synthetic phase distinctions~\cite{niu-sterling-grodin-harper>2022,grodin-li-harper>2025}. 

\begin{axiom}\label{ax:syntax}
    Syntactically, there exists syntax $\red{\tp}$, $\red{\tm}$, $\red{\bool}$, $\red{\true}$, $\red{\false}$, $\red{\ifelim}$, $\red{\pitp}$, $\red{\lam}$, $\red{\app}$, \etc
\end{axiom}

\Cref{ax:syntax} axiomatizes the syntax of the object type theory using higher-order syntax. Most notably, the syntax is \emph{not} inductively defined; in fact, the entire synthetic development of STC avoids any inductive definitions. Therefore, internally, the STC proof cannot perform case analysis on the syntax. This axiom is essential for the concise presentation of syntax in STC, as the use of higher-order abstract syntax simplifies bindings and substitutions significantly compared to first-order representations. 

It is then a natural question to ask: where is the induction proof in STC? The answer is that the induction proof is \emph{analytical} with respect to the \emph{synthetic} development of STC, \ie in the categorical interpretation as shown in \cref{sec:gluing-overview}. The category that represents the syntax, $\T$, is inductively constructed as the free LCCC generated by the signature of \cref{ax:syntax}, and the functorial semantics $\iota : \pre{\T} \to \G$ is defined by induction on this inductively defined category. In other words, the synthetic proofs $\blue{\TP}$, $\blue{\TM}$, $\blue{\BOOL}$, \etc are pieces of lemmas that can be used to construct the inductive, analytical proof. 

The advantage of this approach is the level of abstraction: the analytical justification of \Cref{ax:phase-classifier} and \cref{ax:syntax} in terms of category theory only needs to be carried out once for all object languages, and only the type-theoretic, synthetic proofs need to be developed for each case study. The mechanization of this work, therefore, focuses exclusively on the synthetic part of STC, namely the core proofs of the logical relations argument.

\section{The \istari Proof Assistant}\label{sec:istari}

\istari{}~\cite{istarilogic} is a recently developed, tactic-oriented proof assistant based on Martin-L\"{o}f extensional type theory~\cite{martin-lof>1982}, following the tradition of \lcf~\cite{milner>1972} and \nuprl{}~\cite{nuprl>1986}. It represents a significant extension of traditional \nuprl-style proof assistants, providing support for guarded recursion, impredicativity, and other features. \istari provides a \coq-style user experience to build proof scripts using tactics. This section provides a brief overview of \istari{} and its type theory, focusing on aspects most relevant to the present mechanization.

In \istari terms and computations exist prior to their typing. Types in \istari{} represent partial equivalence relations (PERs) on terms. The judgment $\Gamma \vdash \istEq<A>{M}{N}$ asserts that the terms $M$ and $N$ are equal as elements of type $A$. The derived typing judgment $\Gamma \vdash \istTyOf{M}{A}$ indicates that $M$ is a reflexive instance of the equality $\Gamma \vdash \istEq<A>{M}{M}$. The logic of \istari is a \emph{type-assignment system}~\cite{istarilogic}, allowing users to assign types to terms through automatic and manual typing proofs. Some terms may be assigned multiple types, and others may not be assigned any type at all. Similarly, two terms may be equal at one type but distinct at another.

\istari{} is an \emph{extensional} type theory, in which \emph{judgmental} and \emph{propositional} equality coincide, through the principle of \emph{equality reflection}~\cite{martin-lof>1982,martin-lof>1984}. Therefore, notationally we write $\istEq<A>{M}{N}$ for both judgmental equality and propositional equality. Consequently, a typing judgment $\istTyOf{M}{A}$, being a reflexive instance of equality, also constitutes a proposition within the logic. Typing proofs may involve any mathematical facts, including previously established equations; as a result, type-checking is undecidable in general. In practice, the included type-checker uses known and previously proved typing relations to discharge most of the type-checking goals.

\emph{Computationally equal}~\cite{istarilogic} terms may be converted to one another in a type-free manner. For instance, to establish $\istTyOf{(\lambda\,x.M)~N}{A}$, it suffices to show for the $\beta$-reduct $\istTyOf{M[N/x]}{A}$. Computational equality is closely related to {direct computation} in \nuprl{}~\cite{nuprl>1986,howe>1989,allen-bickford-constable-eaton-kreitz-lorigo-moran>2006}.
This principle aligns with the computational interpretation of terms and reflects the philosophy of meaning explanations from Martin-L\"{o}f type theory~\cite{martin-lof>1982}. 

Equalities in \istari are dictated by the type, similar to those of observational type theory~\cite{altenkirch-mcbride-swierstra>2007,pujet-tabareau>2022}. Conceptually, \istari terms are programs corresponding to the computational content of proofs. Types classify the computational behavior of terms, and computationally equal terms are equal at that type. Concretely, \istari supports:
\begin{enumerate}[leftmargin=*,nosep]
    \item \emph{Function extensionality}: two functions $F$ and $G$ are equal ($\istEq<A \to B>{F}{G}$) if and only if $\istEq<A>{M}{N}$ implies $\istEq<B>{F\,M}{G\,N}$.
    \item \emph{Uniqueness of identity proofs}: any proof of equality $\istEq<A>{M}{N}$ is equal to the trivial empty tuple $\istTriv$.
\end{enumerate}
Equality at universes $\istUniv\,i$ is \emph{intensional}: for example, product types are equal $\istEq<\istUniv\;i>{A \times B}{A' \times B'}$ if and only if their components are equal $\istEq<\istUniv\;i>{A}{A'}$ and $\istEq<\istUniv\;i>{B}{B'}$.

\istari supports a wide range of types beyond what is typically available in proof assistants based on dependent type theory.
Most relevant to this work, \istari supports:
\begin{enumerate}[leftmargin=*,nosep]
    \item \emph{Subtyping.} A type $A$ is a \emph{subtype} of $B$, written $\istSubTy{A}{B}$, if equality at type $A$ implies equality at type $B$. Consequently, if $\istTyOf{M}{A}$, then $\istTyOf{M}{B}$. Subtyping is reflexive and transitive. 
    \item \emph{Cumulative universes.} \istari{} supports a hierarchy of cumulative universes $\istUniv\,0, \istUniv\,1, \istUniv\,2, \ldots$, where $\istSubTy{\istUniv\,i}{\istUniv\,{(1+i)}}$ for every $i$.
    \item \emph{Intersection types.} A term $M$ inhabits the intersection type $\istInter{A}{x}{B}$ if and only if, for every term $N$ such that $\istTyOf{N}{A}$, $\istTyOf{M}{B(N)}$ holds. In other words, the same term $M$ inhabits every member of the type family $B$. Intersection types are particularly convenient for handling universe levels. For instance, $\istTyOf{A}{\istInter{\istLevel}{i}{\istUniv\,i}}$ states that $A$ is a type at any universe $i$.
    \item \emph{Subset types.} \istari{} provides support for working with proof irrelevance with a range of tools. A term $M$ inhabits the subset type $\istSubset{A}{x}{P}$ if $\istTyOf{M}{A}$ and $P(M)$ is inhabited. The proof is \emph{irrelevant} for the equality between elements of the subset type; in particular, $\istSubTy{\istSubset{A}{x}{P}}{A}$.
    \item \emph{Guarded types.} Another tool for proof irrelevance is the guarded function type $A \overset{g}{\to} B$. To establish $\istTyOf{M}{\istGuard{A}{B}}$, it is sufficient to derive $\istTyOf{M}{B}$, with $\istTyOf{x}{A}$ as a ``proof-irrelevant'' assumption. Dually, to use $\istTyOf{M}{\istGuard{A}{B}}$, it suffices to use it as $\istTyOf{M}{{B}}$ and to establish $A$ using tactics. Guarded types serve to encode \emph{presuppositions} such as those in the strict glue types.
\end{enumerate}

The meta-theory of \istari, including soundness and consistency, has been mechanized in \coq~\cite{crary>istari}. The remainder of this section presents a few simple examples, both to elaborate on the preceding discussion and to illustrate additional characteristics of the system.

\subsection{\istari by Example}

Define $\vecTy<A>{n}$ to be the type of $n$-element vectors of type $A$. For the purpose of demonstration, $\vecTy<A>{n}$ is specified as the subset of lists whose computed length is $n$.
\[
    \vecTy<A>{n} \triangleq \istSubset{\listTy<A>}{x}{\istEq{\istLength (x)}{n}{\kw{nat}}}
\]

Following the \lcf tradition, the \istari proof system consists of a trusted kernel and an interface. 
The kernel maintains the current proof state, and the interface allows users to manipulate proof objects 
by invoking tactics and effectful functions on the kernel. For instance, the $\kw{vec}$ type can be defined in \istari as follows:

\iblock{
    \mrow{\istDefine{\kw{vec}~A~n}}
    \mrow{\istParsed{\{ x : \kw{list}~A \mid \kw{length}(x) = n : \kw{nat}\}}}
    \mrow{\istParsed*{\Label{intersect}~i~.~ \Label{forall}~ (A : \Label{U}~i) ~(n : \kw{nat})~.~\Label{U}~i}}
}

The command ``$\Label{define}$'' introduces a new definition, specified by its name, parameters, raw term, and type. 
The type $\vecTy<A>{n}$ is made universe polymorphic by introducing a universe level $i$ through intersection; 
consequently, $i$ does not appear among the parameters. 

In the proof mode the goal is to prove the declared typing. This is discharged with three tactics.
\iblock{
    \mrow{\Label{inference}.~ \Label{unfold}~/\kw{vec}/.~ \Label{typecheck}.}}

\noindent The $\Label{inference}$ tactic performs unification and fills in implicit arguments, such as the type of the variable $i$. 
The defined constant is then unfolded, and finally the type-checker resolves all remaining goals.

As a first example, consider the operation of appending two vectors $v_1$ and $v_2$.

\iblock{
    \mrow{\istDefine{\kw{append}~\{A\}~v_1~v_2}}
    \mrow{\istParsed{\kw{List}.\kw{append}~v_1~v_2}}
    \mhang{\istParsedBegin{\Label{intersect}~i~m~n~.~ \Label{forall}~ (A : \Label{U}~i)}}{
        \mrow{\istParsedEnd*{(v_1 : \kw{vec}~A~{m}) ~(v_2 : \kw{vec}~A~{n}) ~.~ \kw{vec}~A~{(m + n)}}}
    }
}

The function takes in two vectors $v_1$ and $v_2$ of lengths $m$ and $n$, respectively. Its definition is simply $\kw{List}.\kw{append}$, exactly the computation required to append two 
vectors, without any explicit reasoning about lengths. The length constraints are instead handled in the typing proof. The typing proof begins by destructing $v_1$ and $v_2$, which produces four assumptions and a new proof goal:

\iblock{
    \mrow{v_1~v_2 : \kw{list}~A}
    \mrow{H_1~(\Label{hidden}) : \kw{List}.\kw{length}~v_1 = m : \kw{nat}}
    \mrow{H_2~(\Label{hidden}) : \kw{List}.\kw{length}~v_2 = n : \kw{nat}}
    \mhang{\vdash~\kw{List}.\kw{append}~v_1~v_2 }{
        \mrow{: \{ x : \kw{list}~A \mid \kw{List}.\kw{length}~(x) = m + n : \kw{nat}\}}
    }
}

In addition to $v_1, v_2 : \listTy<A>$, two \emph{hidden} assumptions about their lengths are generated. A hidden assumption can only be used in proof-irrelevant proof goals, such as typing proofs, as is the case here.   

The next tactic $\Label{splitOf}$ establishes inhabitation of a subset type by requiring two proofs: first, that the result is a $\listTy<A>$; and second, that its length equals the sum of the lengths of the arguments. The first obligation is automatically discharged by the $\Label{auto}$ tactic. The second requires a lemma from the $\kw{List}$ library. A complete chain of tactics for this interaction is shown below.

\iblock{
    \mrow{\Label{inference}. ~ \Label{unfold}~/\kw{append},~\kw{vec}~\Label{at}~\Label{all}/.}
    \mrow{\Label{introOf}~/i~m~n~A~v_1~v_2/.}
    \mrow{\Label{destruct}~/v_1/~/v_1~H_1/. ~ \Label{destruct}~/v_2/~/v_2~H_2/.}
    \mrow{\Label{unhide}.}
    \mrow{\Label{splitOf}~\text{>>}~\Label{auto}.}
    \mrow{\Label{subst}~/m~n/.}
    \mrow{\Label{apply}~/\kw{List}.\kw{length\_append}/.}
}
As illustrated by this example, \istari provides a streamlined process with a clear separation between the computational content ($\kw{List}.\kw{append}$) and the corresponding correctness argument via subset types.

\subsection{Equality Reasoning in \istari}

\istari offers a variety of tools to alleviate the difficulties of equality reasoning, most notably through equality reflection. As a result, terms often remain close to their intended computational intuition. This section illustrates several of these tools in \istari, with emphasis on techniques used in the mechanization of STC.

\subsubsection{Extensionality.} Heterogeneous equality can be approached directly through extensionality from either side. 
Consider the associativity of $\kw{append}$: for vectors $v_1, v_2, v_3$, 
$$
\kw{append}\, (\kw{append}\,v_1\, v_2)\, v_3 \;=\; 
\kw{append}\, v_1\, (\kw{append}\,v_2\, v_3).
$$
In an intensional type theory, such a statement is not immediately well-typed because the two sides have heterogeneous types and thus one  must transport at least one side. By contrast, \istari permits such heterogeneous equalities as-is.

\iblock{
    \mrow{\Label{lemma}~\kw{assoc}}
    \mhang{\istParsedBegin{\Label{forall}~ i~ (A : \Label{U}~i)~ n_1~ n_2~ n_3~}}{
        \mrow{ (v_1 : \kw{vec}~A~n_1)~ (v_2 : \kw{vec}~A~n_2)~ (v_3 : \kw{vec}~A~n_3) ~.~}
        \mrow{\kw{append}~(\kw{append}~v_1~v_2)~v_3 = \kw{append}~v_1~(\kw{append}~v_2~v_3) : \_} \hfill /;
    }
}
To complete this proof, it suffices to appeal to associativity of $\kw{List}.\kw{append}$ and use tactics to reason about underlying equalities induced by subset types.

\subsubsection{Origami with $\Label{fold}$ and $\Label{unfold}$}\label{sec:istari:origami}
The final example illustrates a technique extensively used in this work to manage coercions via the identity function. 
This technique, which we call \emph{Origami}, resembles the use of transport in intensional type theory, but differs fundamentally in that it relies on the computational content of the identity function. Consider the definition of a reverse function on vectors:
\[
\kw{reverse} : \vecTy{n} \to \vecTy{n} \triangleq \kw{List}.\kw{reverse}.
\]
Now suppose the goal is to establish
\begin{align*}
    &~ \kw{reverse}~(\kw{append}\, v_1\, (\kw{append}\,v_2\, v_3)) \\
    = &~ \kw{reverse}~(\kw{append}\, (\kw{append}\,v_1\, v_2)\, v_3).
\end{align*}
One possible attempt is to directly apply the $\kw{assoc}$ lemma via the $\Label{rewrite}$ tactic, which replaces one side of an equality with the other. However, invoking the tactic directly confuses the type-checker and generates impossible proof obligations, such as $n_1 = n_1 + n_2$. One remedy is to first \emph{fold} a coercion onto one side of the equation along the identity function, and later \emph{unfold} it. In \istari, the coercion along $H : (\istEq{A}{B}{\kw{U}~i})$ can be defined as the following identity function: 

\iblock{
    \mrow{\istDefine{\kw{coe}~H}}
    \mrow{\istParsed{\Label{fn}~ a~ .~ a}}
    \mrow{\istParsed*{\Label{intersect}~ i~(A~B : \Label{U}~i) ~.~ A = B : \Label{U}~i \to A \to B}}
}

Starting with the proof obligation 

\iblock{
    \mrow{\vdash~\kw{reverse}~(\kw{append}~v_1~(\kw{append}~v_2~v_3)) = \kw{reverse}~(\kw{append}~(\kw{append}~v_1~v_2)~v_3) : \_}
}

\noindent the first step is to fold the coercion $\istCoe<H>$ around the right-hand side using the tactic
$\Label{fold}~/\kw{coe}~H/$, producing a homogeneous equality:

\iblock{
    \mrow{H : \kw{vec}~A~((n_1 + n_2) + n_3) = \kw{vec}~A~(n_1 + (n_2 + n_3)) : \Label{U}~i}
    \mrow{\vdash~\kw{reverse}~(\kw{append}~v_1~(\kw{append}~v_2~v_3)) = \kw{reverse}~(\kw{coe}~H~(\kw{append}~(\kw{append}~v_1~v_2)~v_3)) : \_}
}

After this, $\Label{rewrite}$ tactic can be applied without confusion, which results in:

\iblock{
    \mrow{H : \kw{vec}~A~((n_1 + n_2) + n_3) = \kw{vec}~A~(n_1 + (n_2 + n_3)) : \Label{U}~i}
    \mrow{\vdash~\kw{reverse}~(\kw{append}~v_1~(\kw{append}~v_2~v_3)) = \kw{reverse}~(\kw{coe}~H~(\kw{append}~v_1~(\kw{append}~v_2~v_3))) : \_}
}

The key distinction of coercion in \istari compared to its intensional analogue is the ability to $\Label{unfold}~\kw{coe}$, which evaluates the identity function. By contrast, coercions in intensional type theory $\kw{coe} : {(\istEq{A}{B}{\kw{U}~i})} \to A \to B$ do not become the identity function unless the equality in question is reflexivity, which almost always is not the case in a large proof. Unfolding reduces the goal to a reflexive instance, which can be discharged with the $\Label{reflexivity}$ tactic:

\iblock{
    \mrow{\vdash~\kw{reverse}~(\kw{append}~v_1~(\kw{append}~v_2~v_3)) = \kw{reverse}~(\kw{append}~v_1~(\kw{append}~v_2~v_3)) : \_}
}

This technique is extensively applied in this work to facilitate type-checking during rewriting.

\section{Mechanization}\label{sec:mech}
The mechanization of synthetic Tait computability in \istari{} begins with a library for synthetic phase distinction. This library allows on-paper definitions to be transcribed directly into the formalization, with type-checking generating exactly the expected proof obligations. These obligations are then discharged using tactics in \istari. 

\subsection{Library for Synthetic Phase Distinctions}
Because \istari does not have a phase distinction built in, it is extended with a library of definitions and lemmas for phase, modalities, extension types, and strict glue types. This extension axiomatizes the relevant constants and equations as explained in \cref{sec:synthetic}, following the style of logical frameworks~\cite{harper-honsell-plotkin>1993,harper>2021}. In the mechanization, axioms are variables collected in a context. A subset of representative definitions is summarized in this subsection.

\subsubsection{Phase}
A phase is represented as a type \syn{} with at most one element up to equality in \istari.
\iblock{
    \mrow{\kw{syn} : \Label{U}~i}
    \mrow{\kw{syn\_prop} : \Label{forall}~ (z~w : \kw{syn}) ~.~ z = w : \kw{syn}}
}

\subsubsection{Closed Modality}
The closed modality $\closed$ includes two constructors, $\eta_{\closed{}}$ and $\star$, along with an equation $\mathsf{law}$ that identifies them at the syntactic phase.

\iblock{
    \mrow{\kw{closed} : \Label{U}~i \to \Label{U}~i}
    \mrow{\kw{eta} : A \to \kw{closed}~ A}
    \mrow{\kw{star} : \kw{syn} \to \kw{closed}~ A}
    \mrow{\kw{law} : \Label{forall}~ (a : A)~ (z : \kw{syn})~.~ \kw{eta}~a = \kw{star}~z : \kw{closed}~ A}
}

As a pushout/quotient, the closed modality has an eliminator of the following type:

\iblock{
    \mhang{\kw{closed\_elim} : \Label{forall}~ (C : \kw{closed}~ A \to \Label{U}~i)}{
       \mrow{(a : \kw{closed}~ A)}
       \mrow{(c_\eta  : \Label{forall}~ (a : A) ~.~ C(\kw{eta}~a))}
       \mrow{(c_\star : \Label{forall}~ (z : \kw{syn})~.~ C(\kw{star}~z)) ~.~}
       \mrow{(eq    : \Label{forall}~ (a : A)~ (z : \kw{syn}) ~.~c_\eta~a = c_\star ~z : \_) ~\overset{g}{\to} C~a}
    }
}
Most notably, it is not a priori true that this eliminator is well-defined, because $c_\eta~a$ and $c_\star~ z$ have disparate types. Only via the equation $\kw{law}$ can they be identified. In intensional proof assistants such as \agda{}, this term cannot be expressed directly; one must transport one side of the equation along $\kw{law}$. In \istari, the eliminator can be written literally as above because, at that stage, it is merely a raw term. The type-checker generates the proof obligation requesting identification of the types, which is discharged by citing the equation $\kw{law}$ using tactics. This pattern is common when postulating quotient types in proof assistants. Cubical type theories~\cite{angiuli-brunerie-coquand-harper-hou-licata>2021,cohen-coquand-huber-mortberg>2018} can also handle this situation in a computationally well-behaved manner using heterogeneous path types, but equality reflection offers an even simpler solution in situations like this.

Another notable aspect of this definition is the use of the guarded function arrow $\overset{g}{\to}$ in the equality condition $eq$. As introduced in \cref{sec:istari}, guarded functions allow one to avoid explicit equality reasoning in the term; such facts can instead be established in the typing derivation using tactics. This approach lets terms that use the eliminator be written more naturally, as in the definition of $\blue{\IF}$ in \cref{sec:stc}. Using these definitions, lemmas such as the fact that the $\closed$ modality is closed-modal can then be proved.

\subsubsection{Strict Glue Type}
Strict glue types as shown in \cref{fig:sglue-rules} are similarly encoded in \istari. The following is a representative selection of definitions.

\iblock{
    \mhang{\kw{glue\_type} : \Label{forall}~(A : \kw{syn} \to \Label{U}~i)~.~ }{
        \mrow{\Label{forall}~(B : (\Label{forall}~(z : \kw{syn})~.~ A~z) \to \Label{U}~i) ~.~}
        \mrow{(\Label{forall}~a~.~ \kw{closed\_model}~(B~a)) \overset{g}{\to} \Label{U}~i}
    }
    \mrow{}
    \mrow{\kw{glue} : \Label{forall}~(a : \Label{forall}~(z : \kw{syn})~.~ A~z)~.~ (B~a) \to \kw{glue\_type}~A~B}
    \mrow{}
    \mrow{\kw{pi\_open} : \kw{glue\_type}~A~B \to (\Label{forall}~(z : \kw{syn})~.~ A~z)}
    \mrow{\kw{pi\_closed} : \Label{forall}~(g : \kw{glue\_type}~A~B)~.~ B~(\kw{pi\_open}~g)}
    \mrow{}
    \mhang{\kw{type\_eq\_syn} : \Label{forall}~(z : \kw{syn})~.~ }{
        \mrow{\kw{glue\_type}~A~B = A~z : \Label{U}~i}
    }
    \mhang{\kw{term\_eq\_syn} : \Label{forall}~(z : \kw{syn})~(g : \kw{glue\_type}~A~B)~.~ }{
        \mrow{(\kw{pi\_open}~g)~z = g : A~z}
    }
}

\subsubsection{Extension Type}\label{sec:mech:ext}
Extension types $\extTy{A}{\syn{}}{a_0}$ are implemented as subset types as follows:

\iblock{
    \mrow{\istDefine{\kw{ext}~A~a_0}}
    \mrow{\istParsed{\{ x : A \mid \Label{forall}~ (z : \kw{syn}) ~.~ x = a_0~z : A\}}}
    \mrow{\istParsed*{\Label{forall}~ (A : \Label{U}~i) ~.~ (\kw{syn} \to A) \to \Label{U}~i}}
}

The most useful lemma about extension types is that $\extTy{A}{\syn{}}{a_0}$ is a subtype of $A$, which is extensively used in our development to prove that if $a : \kw{ext}~A~a_0$ then $a : A$. This lemma follows directly from subset types.

\iblock{
    \mrow{\Label{lemma}~\kw{ext\_subtype}}
    \mrow{\istParsed*{\Label{forall}~ (A : \Label{U}~i)~ (a_0 : \kw{syn} \to A) ~.~ \kw{ext}~{A}~{a_0} <: A}}
}

\subsection{Definitions of STC in \istari}
The remainder of the development consists largely of a straightforward transcription of the on-paper definitions from \cref{sec:stc} into \istari, using tactics to prove each term has the correct type. This process is mechanical and raises few surprises, highlighting the effectiveness of \istari for mechanizing synthetic Tait computability. As expected, all proof obligations required by STC arise naturally and automatically as type-checking goals in \istari, and are then discharged using tactics in a mostly straightforward manner. 

This process can be illustrated using the definition of $\blue{\LAM}$ from \cref{sec:stc}. As a reminder, the definition is:

\iblock{
    \mrow{\Label{\blue{\LAM}} : \extTy{((x : \blue{\TM}(A)) \to \blue{\TM}(B(x))) \to \blue{\TM}(\blue{\PITP}~A~B)}{\syn}{\red{\lam}}}
    \mrow{\Label{\blue{\LAM}}~f = \glueEx{\syn{}}{\red{\lam}~f}{f} }
}

\noindent That is, the semantics of a lambda function whose type is $\blue{\TM}(\blue{\PITP}~A~B)$ is a syntactic lambda $\red{\lam}~f$ and a semantic function space itself $f : ((x : \blue{\TM}(A)) \to \blue{\TM}(B(x)))$. This definition is expressed in \istari as follows:

\iblock{
    \mrow{\istDefine{\kw{LAM}~A~B}}
    \mrow{\istParsed{\Label{fn}~ f~ .~ \kw{glue}~(\Label{fn}~ z~ .~ \kw{lam}~f)~f}}
    \mhang{\istParsedBegin{\Label{forall}~ (A : \kw{TP})~(B : \kw{TM}(A) \to \kw{TP})~.}}{
        \mrow{\kw{ext}~((\Label{forall}~(x : \kw{TM}(A)).~ \kw{TM}(B(x))) \to \kw{TM}(\kw{PI}~A~B))~(\Label{fn}~ z~ .~ \kw{lam})} \hfill/;
    }
}

\noindent The type-checker in \istari generates the following proof obligations for this definition:

\iblock{
    \mrow{z : \kw{syn}}
    \mhang{\vdash~(\Label{fn}~ f ~.~ \kw{glue}~(\Label{fn}~ z~ .~ \kw{lam}~f)~f) = \kw{lam}}{
        \mrow{: ((\Label{forall}~(a : \kw{tm}(A))~.~ \kw{tm}(B(a))) \to \kw{tm}(\kw{pi}~A~B))}
    }
}

\noindent This condition, under the syntactic phase $\blue{\LAM}$ is equal to $\red{\lam}$, emerges as a type-checking goal when the type-checker reaches the rules for the extension type. This corresponds exactly to one of the critical conditions expected from the definition of STC and the gluing argument in general. The goal is advanced using the function extensionality tactic, yielding:

\iblock{
    \mrow{z : \kw{syn}}
    \mrow{f : (\Label{forall}~(a : \kw{tm}(A))~.~ \kw{tm}(B(a)))}
    \mrow{\vdash~\kw{glue}~(\Label{fn}~ z~ .~ \kw{lam}~f)~f = \kw{lam}~f : \kw{tm}(\kw{pi}~A~B)}
}

\noindent The proof is concluded by citing the corresponding equation from the definition of the strict glue type. A further proof obligation arises from the extension type in the definition of $\blue{\PITP}$: it must hold that $\red{\app}~(\red{\lam}~f) = f$, since every term of type $\red{\pitp}~A~B$ is characterized by its application $\red{\app}$. This appears as

\iblock{
    \mrow{z : \kw{syn}}
    \mrow{f : (\Label{forall}~(a : \kw{tm}(A))~.~ \kw{tm}(B(a)))}
    \mrow{\vdash~\kw{app}~(\kw{lam}~f) = f : (\Label{forall}~(a : \kw{tm}(A))~.~ \kw{tm}(B(a)))}
}

\noindent This goal is discharged by citing the corresponding equation $\red{\pitp_\beta}$ from syntax.

For each constant in the STC definition, the same process is followed: the on-paper term is transcribed verbatim into \istari and then type-checked using tactics. The default type-checker of \istari automatically discharges most proof obligations, and the remaining ones typically capture the essential content of the STC proof, the parts that would otherwise be justified informally in an on-paper development. These are then handled manually by citing the relevant equations and lemmas. As one might expect, extensional type theories such as \istari enable all definitions to be expressed essentially verbatim without modification, thereby minimizing the gap between the on-paper and formalized proofs.

\subsection{Proof Engineering in \istari}
This subsection discusses techniques used to streamline equality reasoning in \istari with concrete examples from the mechanization of STC.

\subsubsection{Hinting the Type-checker with Origami}
A notable phenomenon of synthetic phase distinction is that a term may inhabit different types depending on the phase. For instance, a term of type $\blue{\TP}$ is also of type $\red{\tp}$ under the syntactic phase, by virtue of the identification $\open(\blue{\TP} = \red{\tp})$. Consider, for example, the definition of $\blue{\PITP}$:

\iblock{
    \mrow{\Label{\blue{\PITP}} : \extTy{(A : \blue{\TP}) \to (\blue{\TM}(A) \to \blue{\TP}) \to \blue{\TP}}{\syn}{\red{\pitp}}}
    \mrow{\Label{\blue{\PITP}}~A~B = [ \syn{} \hookrightarrow \red{\pitp}~A~B \mid \cdots ]}
}

\noindent To type-check the syntactic part of this definition, \istari will generate the following proof obligation:

\iblock{
    \mrow{z : \kw{syn}}
    \mrow{A : \kw{TP}}
    \mrow{B : \kw{TM}(A) \to \kw{TP}}
    \mrow{\vdash~\kw{pi}~A~B : \kw{tp}}
}

\noindent Because $\kw{pi} : \Label{forall}~(A : \kw{tp})~.~ (\kw{tm}(A) \to \kw{tp}) \to \kw{tp}$ and $A : \kw{TP}$, the type-checker will further generate the sub-goal:

\iblock{
    \mrow{z : \kw{syn} \vdash~\kw{TP} = \kw{tp} : \Label{U}~i}
}
\noindent which can then be proved using the extension type property on $\kw{TP}$. 
However, this proof obligation shows up often enough that it became tedious to discharge it manually. 
With the Origami technique described in \cref{sec:istari:origami} we can streamline it by inserting ``type casts'' that 
point the type-checker to a specific type equality, without affecting the extent of the term itself.
To this end, define 
$$
\kw{cast}~z \triangleq \kw{coe}\,(H\,z), \quad\text{where}\quad
H : \open(\blue{\TP} = \red{\tp}).
$$

\noindent Definition of $\kw{PI}$ can then be modified to impose $\kw{cast}$ on $A$:

\iblock{
    \mrow{\istDefine{\kw{PI'}~A~B}}
    \mrow{\istParsed{\Label{fn}~A~B~ .~ \kw{glue}~(\Label{fn}~z~.~ \kw{pi}~(\kw{cast}~z~A)~\cdots)~\cdots}}
    \mhang{\istParsedBegin{{\Label{forall}~ (A : \kw{TP})~(B : \kw{TM}(A) \to \kw{TP}) .~}}}{
        \mrow{\kw{ext}~((\Label{forall}~(x : \kw{TM}(A))~.~ \kw{TM}(B(x))) \to \kw{TM}(\kw{PI}~A~B))~(\Label{fn}~ z~ .~ \kw{pi})} \hfill/;
    }
}

\noindent The proof obligation $\kw{TP} = \kw{tp}$ no longer arises during type-checking, as the type of $\kw{cast}~z~A$ is already $\kw{tp}$. 
As in \Cref{sec:istari:origami}, this ``type cast'' is computationally just an identity function. It can subsequently be eliminated by $\Label{unfold}$ing relevant definitions, yielding

\iblock{
    \mhang{(\Label{fn}~A~B~ .~ \kw{glue}~(\Label{fn}~z~.~ \kw{pi}~A~\cdots)~\cdots)}{
        \mrow{: \Label{forall}~ (A : \kw{TP})~(B : \kw{TM}(A) \to \kw{TP}) ~.~ \cdots.} 
    }
}

\noindent This recovers exactly the original on-paper definition of $\blue{\PITP}$. Notably, the situation differs from coercions in intensional proof assistants, where computation of a coercion is possible only when the underlying equality is reflexivity, a condition rarely satisfied within a large proof. The definition of $\kw{cast}$ plays an additional role in the development: proof obligations of the following form often arise after converting $\kw{TM}$ to $\kw{tm}$:

\iblock{
    \mrow{z : \kw{syn}}
    \mrow{A : \kw{TP}}
    \mrow{\vdash~\kw{tm}~A : \Label{U}~i}
}

\noindent This induces the familiar sub-goal $\kw{TP} = \kw{tp}$ as before. In order to let the type-checker discharge this goal, 
the opposite strategy from \Cref{sec:istari:origami} can be used: rather than eliminating a coercion from a definition, 
one introduces it within a proof by imposing $\kw{cast}$ on $A$ via the $\Label{fold}$ tactic.

\iblock{
    \mrow{\Label{fold}~/\kw{cast}~z~A/.}
}

\noindent The $\Label{fold}$ tactic changes the goal to
$\kw{tm}~(\kw{cast}~z~A) : \Label{U}~i$,
which the type-checker discharges automatically. This proof engineering technique becomes particularly significant in the presence of extension types. For instance, $\kw{TM}(A)$ is frequently used to denote the type of terms of type $A$. The type-checker has difficulty with this, because the type of $\kw{TM}$ is an extension type, yet it is applied as if it were a function with argument $A$.

\iblock{
    \mrow{\kw{TM} : \kw{ext}~(\kw{TP} \to U~i)~(\Label{fn}~ z~ .~ \kw{tm})}
}

\noindent The type-checker tries to unify the extension type and a function type, creating an impossible goal:

\iblock{
    \mrow{\vdash~(\kw{TP} \to U~i) =  \kw{ext}~(\kw{TP} \to U~i)~(\Label{fn}~ z~ .~ \kw{tm}) : \Label{U}~(1 + i)}
}

\noindent As before, Origami can guide the type-checker to coerce between extension types and their original types. $\kw{coe}~H$ in 
\cref{sec:istari:origami} can be similarly extended to enable passage from subtypes to super-types when $H$ proves a subtyping. Citing 
$\kw{ext\_subtype}$ from \cref{sec:mech:ext}, we define
$$\kw{out} \triangleq \kw{coe}~\kw{ext\_subtype}.$$

\noindent The problematic $\kw{TM}(A)$ is instead written as $(\kw{out}~\kw{TM})~A$, which type-checks without issue. $\kw{out}$ can be eliminated as before, recovering the on-paper definitions. This behavior contrasts with formulations of extension types in which $\kw{in}$ and $\kw{out}$ are primitive introduction and elimination forms~\cite{riehl-shulman>2017,zhang>2024} that cannot be ``computed away'' on their own.\footnote{An analogy for this distinction is the difference between equal-recursive
and iso-recursive types.}

\subsubsection{Computational Equality}
The semantics of large elimination for booleans, $\blue{\IF}$, together with its associated equation $\blue{\IF_{\beta_1}}$, provide a representative example of the usefulness of computational equality in \istari, as discussed in \cref{sec:istari}. Concretely, the semantics of $\blue{\BOOL}$ is given by a binary sum, $\blue{\TRUE}$ corresponds to a left injection, and $\blue{\IF}$ is a case analysis. In verifying the equation
\[
  \blue{\IF_{\beta_1}} : (\blue{\IF}~C~\blue{\TRUE}~t~f = t : \blue{\TM}(C(\blue{\TRUE})))
\]
the following proof obligation arises:

\iblock{
    \mrow{C : \kw{TM}(\kw{BOOL}) \to \kw{TP}}
    \mrow{t : \kw{TM}(C(\kw{TRUE}))}
    \mrow{f : \kw{TM}(C(\kw{FALSE}))}
    \mrow{\vdash~(\Label{case}~(\Label{inl}~())~\Label{of}~\Label{|}~\Label{inl}~\_~\Label{.}~t~\Label{|}~\Label{inr}~\_~\Label{.}~f) : \kw{TM}(C(\kw{TRUE}))}
}

\noindent To type-check this goal directly, the type-checker needs to reason about the impossibility of the second branch. With computational equality, it is possible to first run a type-free computation on the term and then type-check the resulting term. 
Using the $\Label{reduce}$ tactic of \istari,
the term in question is simplified according to the type-free operational semantics:
$$(\Label{case}~(\Label{inl}~())~\Label{of}~\Label{|}~\Label{inl}~\_~\Label{.}~t~\Label{|}~\Label{inr}~\_~\Label{.}~f) \mapsto t.$$ 
The new goal $t : \kw{TM}(C(\kw{TRUE}))$ is immediate.

\subsection{Case Studies}
The effectiveness of this formalization is demonstrated by two case studies of STC applications.

\subsubsection*{Core Dependent Type Theory}
The first case study is the canonicity gluing model of a dependent type theory with dependent product types and booleans with large elimination, exactly as presented in \cref{sec:stc}. Each constant in the proof of STC is transcribed exactly as on-paper from~\citet{sterling>thesis,sterling>2022-logical-relations} into \istari as terms and types, followed by type-checking using tactics. The terms themselves remain as concise as in the on-paper development. 
We report the rough number of lines of tactics needed for type-checking each constant below.

\begin{center}
\begin{tabular}{c c}
    \toprule
    \textbf{Lines of tactics} & \textbf{Definitions} \\
    \midrule
    0--100     & \blue{\TM}, \blue{\TP}, \blue{\TRUE}, \blue{\FALSE} \\
    100--200   & \blue{\PITP}, \blue{\LAM}, $\blue{\PITP_\beta}$, \blue{\BOOL} \\
    400--500   & \blue{\APP}, $\blue{\PITP_\eta}$ \\
    1200--1500 & \blue{\IF}, $\blue{\IF_{\beta_1}}$, $\blue{\IF_{\beta_2}}$ \\
    \bottomrule
\end{tabular}
\end{center}

In some cases, the tactic scripts are longer than expected because the modality framework is axiomatized inside \istari rather than being built in as primitives, so the type-checker does not have as good support for reasoning about, \eg, strict glue types as it would have if they were built in. For example, unlike built-in dependent sums, strict glue types do not enjoy type-free computation, so manual tactics are needed to go through certain steps in $\blue{\PITP_\beta}$. Similarly, having to work with axiomatized large elimination for the $\closed$ modality in $\blue{\IF}$ leads to longer proof scripts. Nevertheless, the overall length of the tactic scripts remains quite manageable, and this only impacts tactic scripts rather than the terms themselves.

\subsubsection*{Cost-Aware Logical Framework}
Cost-aware logical framework (\calf{})~\cite{niu-sterling-grodin-harper>2022,grodin-niu-sterling-harper>2024} is a dependent call-by-push-value language designed for synthetic cost analysis of algorithms~\cite{li-grodin-harper>2023,grodin-harper>2024,zhou>2025,kebuladze>2025}. \calf{} incorporates a phase distinction $\beh$ between cost and behavior, analogous to $\syn$, to isolate the behavioral aspects of a program from its cost. Its call-by-push-value~\cite{levy>2003} structure includes a free-forgetful adjunction $\mathsf{F} \dashv \mathsf{U}$, with an underlying writer monad used for cost tracking, and a cost-charging computational effect $\mstep{c}{e}$ that records $c$ units of cost when executing the computation $e$.

The canonicity property of \calf{} is formally established in~\citet{li-harper>2025}, where a gluing model is developed using STC. \calf{} is justified by a two-world Kripke semantics, \ie presheaves over the poset $\vmathbb{2} = \{\beh \to \top\}$. To be precise, the gluing category $\G$ is taken to be $\pre{\vmathbb{2}} \downarrow N_\rho$ where $N_\rho$ is the global sections functor induced by the phase-separated interpretation $\rho$ of \calf{} into $\pre{\vmathbb{2}}$ as follows:

\vspace{0.4em}
\begin{tabular}{p{3.5cm}p{5cm}}
$\begin{array}{l}
\rho : \vmathbb{2} \to \pre{\T} \\
\rho(\beh) = \Yo(\red{\beh}) \\
\rho(\top) = \Yo(\red{\terminal}_{\T})
\end{array}$
&
$\begin{array}{l}
N_\rho : \pre{\T} \to \pre{\vmathbb{2}} \\
N_\rho(X) = \homset{\rho-}{X} \\
\phantom{match height} 
\end{array}$
\end{tabular}

\noindent Synthetically this two-world Kripke structure is captured by axiomatizing the $\beh$ phase in the \istari mechanization, in a similar manner to the $\syn$ phase. 

The STC proof then captures the Eilenberg-Moore~\cite{eilenberg-moore>1965} categorical structure for call-by-push-value~\cite{levy>2003} where the free functor $\mathsf{F}$ constructs free monad algebras and the forgetful functor $\mathsf{U}$ forgets the algebra structure. Every computation type in call-by-push-value is interpreted as a monad algebra $\alg$ that satisfies the monad laws as well as an equation $\Label{step_{\beh}}$ that erases cost at the behavioral phase. For any computation type $X$, the syntax of \calf{} forms exactly such an algebra $\mathsf{SynAlg}(X)$ under the syntactic phase $\syn$.

\iblock{
    \mhang{\kw{record}~\alg~\kw{where}}{
        \mrow{\Label{Car} : \univ}
        \mrow{\Label{step} : \mathbb{C} \to \Label{Car} \to \Label{Car}}
        \mrow{\Label{step_{0}} : \Label{step}~0~x = x}
        \mrow{\Label{step_{+}} : \Label{step}~c_1~(\Label{step}~c_2~x) = \Label{step}~(c_1 + c_2)~x}
        \mrow{\Label{step_{\beh}} : \open_{\beh}(\Label{step}~c~x = x)}
    }
}

The STC proof for computation types $\red{\tpc}$ is similar to that for types $\red{\tp}$ in \cref{sec:stc-judgments}, with the addition of verifying that every computation type in the gluing model forms a monad algebra, and that they collapse to the syntactic algebra $\mathsf{SynAlg}$ under the syntactic phase. 

\iblock{
    \mrow{\Label{\blue{\TPC}} : \extTy{\univ}{\syn}{\red{\tpc}}}
    \mrow{\Label{\blue{\TPC}} = \glue{X}{\red{\tpc}}{\extTy{\alg}{\syn}{\textsf{SynAlg}(X)}}}
}

The desired canonicity property for \calf{} is that every closed term of boolean free computation is equal to charging certain cost and returning either $\red{\true}$ or $\red{\false}$. 

\begin{theorem}[Canonicity]
For every term $e : \red{\tmc}(\red{\F}(\red{\bool}))$, there exists $c : \mathbb{C}$ such that $e = \red{\step}^c(\red{\ret}(\red{\true}))$ or $e = \red{\step}^c(\red{\ret}(\red{\false}))$.
\end{theorem}

This property is captured in the carrier type of $\blue{\F}$, which forms the free monad algebra. 

\iblock{
    \mrow{\Label{\blue{\F}} : \extTy{\blue{\TPV} \to \blue{\TPC}}{\syn}{\red{\F}}}
    \mrow{\Label{\blue{\F}}(A) = \glueEx{\syn{}}{\red{\F}(A)}{\mathsf{FreeAlg}(A)}}
    \mrow{}
    \mrow{\mathsf{FreeAlg}(A).\Label{Car} : \extTy{\univ}{\syn}{\mathsf{SynAlg}(\red{\F}(A)).\Label{Car}}}
    \mhang{\mathsf{FreeAlg}(A).\Label{Car} = \glue{e}{\red{\tmc}(\red{\F}(A))}{\closed{}~P}}{
        \mhang{\kw{where}}{
            \mrow{P = \Sigma_{a : \blue{\TMV}(A)} \Sigma_{bc : \closed{}_{\beh} \mathbb{C}} ~\mathsf{case}~bc~\mathsf{of}}
            \mrow{\quad \star~(b : \beh) \Rightarrow \open{} (e = \red{\ret}~a)}
            \mrow{\quad \eta_{\closed{}_{\beh}}~(c : \mathbb{C}) \Rightarrow \open{} (e = \red{\step}^c(\red{\ret}~a))}
        }
    }   
}

In other words, the (proof-relevant) predicate $P$ associated with the $\red{\F}$ type is the collection of all values it can return ($a : \blue{\TMV}(A)$) and the cost charged ($bc : \closed{}_{\beh} \mathbb{C}$). The case analysis on $bc$ amounts to case analysis of the Kripke worlds, making sure that the canonicity property holds in both worlds. Instantiating $A$ to be $\blue{\BOOL}$ then yields the desired canonicity property. For more details about the construction of $\blue{\F}$ and other definitions in the STC proof for \calf{}, we refer the reader to~\citet{li-harper>2025} and the accompanying mechanization of this work where we formalize representative definitions, namely the call-by-push-value adjunction $\blue{\F} \dashv \blue{\U}$ and the behavioral/cost phase distinction with effects $\mstep{c}{e}$.

\section{Related Work}\label{sec:related}

This section discusses related work on prior formalizations of gluing, synthetic Tait computability, and logical relations, and on extensional proof assistants comparable to \istari.

\subsection{Formalization of Gluing Argument}
There has been significant recent progress on formalizing gluing arguments. For example, normalization gluing models for the simply-typed $\lambda$ calculus have been mechanized in \cubical{} and \coq{}~\cite{1lab,berry-fiore>2025}. Most notably,~\citet{kaposi-pujet>2025} formalized a canonicity gluing model for a dependent type theory presented as a category with families in \agda, using postulated constructs from observational type theory~\cite{altenkirch-mcbride-swierstra>2007,pujet-tabareau>2022}. This work differs in several respects.

\subsubsection*{Treatment of Syntax and Equalities}
Both works begin with an algebraic signature of the syntax: theirs, a first-order category with families; ours, a higher-order abstract syntax presentation for a generalized algebraic theory. In such presentations, the equations of the syntax are expressed as propositional equalities. Both works are motivated by the observation that, for proof engineering purposes, it is advantageous to make as many of these equations hold judgmentally as possible. The approach~\citet{kaposi-pujet>2025} takes to \emph{strictify} the equations is to instantiate the signature in a particular way, using quotient inductive-inductive types~\cite{kaposi-kovacs-altenkirch>2019} and techniques from strict presheaves~\cite{pedrot>2020}. This construction ensures that all equations of the substitution calculus hold judgmentally, which considerably simplifies the subsequent gluing construction. Some equations, most notably $\beta$ and $\eta$, remain propositional, so a small amount of transports and coercions are still required in their gluing proof. In the present work, the use of higher-order abstract syntax already avoids many equations about substitutions, most notably the naturality conditions. The remaining equations are uniformly turned into judgmental equalities by \istari's equality reflection, thereby avoiding the need for transports and bringing the formalization closer to the on-paper proofs. Consequently, this work does not make the effort to define the inductive syntax in type theory (in fact, the inductive syntax cannot be defined internally due to the use of higher-order abstract syntax), but rather works directly with the abstract signature of the syntax, without relying on initiality. The use of the modality framework does create some overhead in the number of equalities to manage, as explained in \cref{sec:formalization-intro}, but all such equalities are straightforward to handle with equality reflection as well. The drawback of our approach is that, in \citet{kaposi-pujet>2025}, judgmentally equal terms compute to the same thing automatically, while in our approach judgmentally equal terms are proved equal by manually invoking relevant equations using tactics. 

\subsubsection*{Computational Content}
Working synthetically exposes both advantages and limitations in the formalization. A primary limitation is that, although the proof is fully constructive, an evaluation algorithm is only external and cannot be extracted \emph{internally} in \istari. This arises from the nature of synthetic Tait computability as explained in \cref{sec:synthetic}. By contrast, the algorithmic computational content of~\citet{kaposi-pujet>2025} is directly extractable internally in \agda. In exchange for this limitation internal language constructs can be flexibly reused across different object languages, which is exactly the \emph{synthetic} advantage of STC. For example, although the gluing categories for the dependent type theory and \calf{} differ, their internal languages share largely the same structures, allowing formalizations to use the same library of modal dependent type theory.

\subsection{Formalization of Synthetic Tait Computability}
The present work is not the first to consider mechanizing synthetic Tait computability. \citet{sterling-harper>2021} anticipated that the main difficulty of the mechanization might lie in the treatment of phase, and suggested that definitional proof-irrelevance~\cite{gilbert-cockx-sozeau-tabareau>2019}, as implemented in \agda{} and \coq{}, could be used to implement the phase \syn{}. The present work shows that this is not the main bottleneck: the fact that \syn{} is propositional is used only once in the entire development, namely to show that $\closed{}$ is closed-modal. \citet{huang>2023} suggested using \cubical{} cofibrations and glue types to simulate phases and the realignment axiom, but the semantics of cubical type theory is distant from the extensional internal language of the gluing category $\G$. Although \cubical{} would allow a neat formulation of the \closed{} modality as a higher inductive type, it is overkill because the gluing requires only set-level mathematics, and this work deliberately tries to avoid pervasive set-truncation and transports.

\subsection{Mechanized Logical Relations}

Besides the gluing construction and its synthetic variants, there is a rich literature on mechanizing logical relations and normalization-by-evaluation arguments, analytically rather than synthetically, for dependent, effectful, and/or linear type theories in \agda~\cite{abel-danielsson-eriksson>2023,abel-ohman-vezzosi>2017,acevedo-weirich>2023,danielsson-favier-kubanek>2026} and \coq~\cite{gregersen-bay-timany-birkedal>2021,wieczorek-biernacki>2018,adjedj-lennon-bertrand-maillard-pedrot-pujet>2024,zhang-simkin-li-yao-balzer>2025,liu-chan-weirich>2025,gollamudi-jacobs-yao-balzer>2025}. These mechanizations typically work with an operational semantics or a reduction system, and define logical relations over them. Such syntactic approaches are more flexible in the sense that they can directly manipulate arbitrary syntactic constructs. By contrast, the gluing approach applies most naturally to type theories given more semantic and categorical presentations, such as those with reduction-free equational theories. It is yet unclear how to apply gluing to languages that are primarily oriented around operational semantics. Another common challenge these mechanizations face is the tedious reasoning about bindings and substitutions, though this can be mitigated by automated tools~\cite{stark-schafer-kaiser>2019,schafer-tebbi-smolka>2015}. The synthetic approach we take here avoids these challenges by using higher-order abstract syntax, in a style similar to mechanizing logical relations in \beluga~\cite{poplmark-reloaded,cave-pientka>2018}.

\subsection{Other Related Proof Assistants}

\istari{}~\cite{istarilogic} follows the computational type theory tradition of Martin-L\"{o}f type theory~\cite{martin-lof>1982} and \nuprl~\cite{nuprl>1986}, whose computational semantics inspired \metaprl{}~\cite{metaprl>2003} and higher-dimensional proof assistants such as \redprl{}~\cite{angiuli-hou-harper>2018,angiuli>thesis,angiuli-cavallo-hou-harper-sterling>2018}. Our development could in principle live in these systems, especially \nuprl, but we use \istari for its modern, stable implementation.

Outside this tradition, several type theories validate extensional principles. Observational type theory (OTT)~\cite{altenkirch-mcbride-swierstra>2007,pujet-tabareau>2022} validates function extensionality but not equality reflection, and its \agda{} embeddings support gluing arguments~\cite{kaposi-pujet>2025}. \citet{sterling-angiuli-gratzer>2022,sterling-angiuli-gratzer>2019} propose a type theory with uniqueness of identity proofs using cubical syntax, currently without an implementation. \andromeda~\cite{bauer-gilbert-haselwarter-pretnar-stone>2018} implements an extensional type theory with equality reflection, but explores different proof-engineering choices than \istari.

\section{Conclusion and Future Work}\label{sec:future}

This work mechanizes synthetic Tait computability in the \istari{} proof assistant, which is based on extensional type theory. Our guiding principle throughout is that computer formalizations should track the on-paper proof as closely as possible, introducing as little extraneous technical machinery as possible. By taking full advantage of \istari's equality reflection and related features, the mechanization remains straightforward and largely preserves the concision and elegance of the original arguments, in line with the vision of~\citet{sterling>thesis}, where STC reduces complex gluing constructions to simple type-theoretic reasoning.

This perspective addresses a common concern in the literature that such arguments are difficult to mechanize:
\begin{myquote}{adjedj-lennon-bertrand-maillard-pedrot-pujet>2024}
Accordingly, these proofs [gluing] have a very extensional flavor,
and as such are less amenable to implementation in a proof
assistant based on intensional type theory. Moreover, the
more sophisticated iterations [STC] rely on (multi)modal type theories as internal languages for feature-rich categories, for
which mechanization is still in its infancy. 
\end{myquote}
Overall, we believe our development demonstrates the feasibility and benefits of mechanizing complex meta-theoretic arguments in an extensional proof assistant. The main price we pay for concise terms that closely resemble on-paper proofs is a certain amount of manual type-checking, a consequence of undecidable type-checking. In practice, however, we find that \istari's tactics and automation alleviate much of this burden to a manageable level. This work also serves as a proof of concept that extensional proof assistants with equality reflection provide a viable and promising alternative to intensional proof assistants for specific classes of applications, such as are demonstrated here.

\subsection{Future Work}
The present mechanization of synthetic Tait computability in \istari{} opens several avenues for future research:

\subsubsection*{Further Formalizations} 
The library and methods in this work should be able to extend to formalizations of more sophisticated STC proofs, such as:
\begin{enumerate}[leftmargin=*,nosep]
    \item binary homogeneous logical relations for parametricity of a module calculus~\cite{sterling-harper>2021};
    \item binary heterogeneous logical relations for compiler, \eg call-by-value to call-by-push-value compilation;
    \item step-indexed logical relations for recursive types~\cite{sterling-gratzer-birkedal>2023}, ready for mechanization in \istari{} thanks to \istari's support for the future modality and guarded recursion;
    \item normalization for dependent type theory~\cite{sterling>thesis,gratzer>2022,sterling-angiuli>2021,gratzer-sterling-angiuli-coquand-birkedal>2025}.
\end{enumerate}

\subsubsection*{Mechanized Analytical Justification of STC}
As mentioned in \cref{sec:synthetic,sec:related}, externalizing the internal language to the gluing category $\G$ is necessary to obtain the algorithmic content of STC proofs. Two possible approaches are:
\begin{enumerate}[leftmargin=*,nosep]
    \item Extend \istari's computational semantics to support a presheaf model, \ie Kripke logical relations for the syntax-semantics phase distinction, so the internal mechanization is directly justified by \istari's semantics;
    \item Mechanize the gluing categorical construction with respect to the internal language of STC in a proof assistant such as \agda{}, \coq{}, or \lean{}, where significant category-theoretic formalizations already exist~\cite{hu-carette>2021,mathlib>2020,unimath>2025}.
\end{enumerate}
 
\section*{Data Availability Statement}
The \istari{} mechanization described in this paper is available at \citet{li-yao-harper>artifact}.

\begin{acks}
The authors thank Karl Crary, Jonathan Sterling, Harrison Grodin, and Matias Scharager for fruitful discussions and proofreading of this work, and the anonymous reviewers for their thoughtful comments. 
This material is based upon work supported by the \grantsponsor{NSF}{National Science Foundation}{https://www.nsf.gov/}
under grant numbers \grantnum{NSF}{2211996} and \grantnum{NSF}{2442461}, and by the \grantsponsor{AFOSR}{United States Air Force Office of Scientific Research}{https://www.afrl.af.mil/AFOSR/} under grant numbers \grantnum{AFOSR}{FA9550-21-0009}, \grantnum{AFOSR}{FA9550-23-1-0434}, and \grantnum{AFOSR}{FA9550-21-1-0385} (Tristan Nguyen, program manager). Any opinions, findings, and conclusions or recommendations expressed in this material are those of the authors and do not necessarily reflect the views of the National Science Foundation or the United States Air Force Office of Scientific Research.
\end{acks}

\bibliographystyle{ACM-Reference-Format}
\balance
\bibliography{main}

\end{document}